\documentclass[11pt,a4paper,twoside]{article}  %
\usepackage[american]{babel}     
\usepackage{amsmath,amsthm,amssymb,latexsym}
\usepackage{amsfonts}           
\usepackage{geometry}
\usepackage{graphicx}         
\usepackage{fancyhdr}
\usepackage{tikz}
\usepackage{color}

\def\RR{\mathbb{R}}

\newenvironment { abstract }

\author{Roberto Franzosi\\
QSTAR \& INO-CNR, largo E. Fermi 2, 50125 Firenze, Italy\\
\and
Domenico Felice\\
School of Science and Technology \\University of Camerino, I-62032 Camerino, Italy\\
INFN-Sezione di Perugia, Via A. Pascoli, I-06123 Perugia, Italy
\and 
Stefano Mancini\\
School of Science and Technology \\University of Camerino, I-62032 Camerino, Italy\\
INFN-Sezione di Perugia, Via A. Pascoli, I-06123 Perugia, Italy
\and
Marco Pettini\\
Centre de Physique Th\'eorique, UMR7332, and Aix-Marseille University, \\ Luminy Case 907, 13288 Marseille, France}

\begin{document}
\title{\bf A geometric entropy detecting the Erd\"os-R\'enyi phase transition}

\maketitle

\begin{abstract}
We propose a method to associate a differentiable Riemannian manifold to a generic many degrees of freedom discrete system which is not described by a Hamiltonian function. Then, in analogy with classical Statistical Mechanics, we introduce an entropy as the logarithm of the volume of the manifold. 
The geometric entropy so defined is able to detect a paradigmatic phase transition occurring in random graphs theory: the appearance of the `giant component' according to the Erd\"os-R\'enyi theorem.
\end{abstract}

\date{}


\section{Introduction}
Thermodynamic phase transitions are examples of emergent phenomena in many degrees of freedom systems, described in the framework of Statistical Mechanics. The standard statistical ensembles measures  relate macroscopic (thermodynamic) observables with microscopic  degrees of freedom. The interactions among the microscopic degrees of freedom - which can be either continuous or discrete (as in the case of spin models, vertex models, and so on) - are { often} described by a Hamiltonian function
(or a Hamiltonian operator in a quantum context) \cite{P07}.  But what about  discrete systems, i.e. networks, undergoing a phase transition for which a microscopic Hamiltonian does not exist? 

A paradigmatic example is represented by a random graphs model $\mathbb{G}(n,k)$ devised by choosing  with uniform probability a graph from the set of all graphs having $N$ vertices and $L$ edges  \cite{Lucz}. We can think of a process evolving by adding the edges one at a time. When $k$ has the same order of magnitude of $n$, the evolution from $k=0$ to $k=\binom{n}{2}$ yields, according to Erd\"{o}s-R\'enyi theorem \cite{ER}, a \textit{ phase transition}, revealing itself  in a rapid growth with $k$ of the size of the largest component (number of vertices fully connected by edges). Specifically, the structure of graphs when the expected degree of each of its vertices is close to $1$, i.e. $k\sim n/2$, shows a jump: the order of magnitude of the size of the largest component of graphs rapidly grows, asymptotically almost surely, from $\log n$ to $n$, if $k$ has the same order of magnitude of $n$.
In fact, if $k<n/2$, as the process evolves, the components of a graph [the largest of them being a.a.s. of size $O(\log n)$] merge mainly by attaching small trees; thus they grow slowly and quite smoothly. Nonetheless, at the same point of the process, the largest components become so large that it is likely for a new edge to connect two of them. Thus, fairly quickly, all the largest components of a graph merge into one giant component, much larger than any of the remaining ones \cite{Lucz}.  
It is worth noticing that this process represents the mean-field case of percolation \cite{PERC}.

Regarding $\mathbb{G}(n,k)$ as a statistical ensemble it is quite natural to employ tools from statistical mechanics, above all entropy, to analyze it. 
In Ref. \cite{BBW06} the Gibbs entropy of such random graphs was defined as  
 \begin{equation}\label{Ssimple}
S:=\ln \frac{1}{n!} \binom{\binom{n}{2}}{k}.
\end{equation}
There, the configuration space was given by $ \binom{\binom{n}{2}}{k}$ graphs with labelled nodes. Due to their equiprobability, they have the same weight, chosen to be $n!$  in order to account for all the labelling permutations of the nodes.  
Later, the perspective was to modify the Erd\"os-R\'enyi ensemble by introducing a functional weight which explicitly depends on the graph's topology. In this way one can eventually characterize other 
classes of random graphs, like scale free or fixed degree sequence, as well.
A research line that has been pursued in the last decade \cite{Bianconi}, also putting forward variants of the entropy measure  \eqref{Ssimple} \cite{ENTROPY}. 
However, we may notice that the entropy \eqref{Ssimple} as a function of the ratio $k/n$ is unable itself to detect the phase transition occurring in the Erd\"os-R\'enyi ensemble.

In the present work, focussing on the  Erd\"os-R\'enyi ensemble, we propose a general method to associate a continuous mathematical object (a Riemannian manifold) to a generic discrete system, graph or network, thus allowing the definition of a \textit{geometric entropy} which is able to detect a phase transition.
Actually, we endow each network with a statistical Riemannian manifold. This can be obtained basically via two steps; first by understanding a network as an undirected graph without loops on the nodes, and account for links (weighted edges) between nodes expressed by the adjacency matrix $A$. Second, considering random variables as sitting on the vertices of a network,  methods of information geometry \cite{AN00} can be  employed to lift the network to a statistical Riemannian manifold. 
In this way, we associate a configuration space to each network. Such a space consists of a subset of the linear vector space $\RR ^m$ given by the parameters which characterize the joint probability distribution of the random variables sitting on the nodes of the network. Furthermore, this configuration space is endowed with a Riemannian metric which is inspired by the well-known Fisher-Rao metric \cite{AN00}.
Then, in analogy with classical Statistical Mechanics, we define a geometric entropy as the logarithm of the volume of this manifold. 
Applied to Erd\"os-R\'enyi random graphs it proves very effective (as a function of the ratio $k/n$) in detecting the appearance of the so called `giant component' as well as any smooth function of order parameters within the framework of Statistical Mechanics \cite{P07}.

%%%%%%%%%%%%%%%%%%%%%%%%%%%%%%%%%%%%%%%%%%%%%%%%%%%%%%%%

\section{Information geometric model}

Let us start considering a set of $n$ real-valued random variables $X_1,\ldots,X_n$ characterized by the following multivariate Gaussian probability distribution  
\begin{equation}
 p(x;\theta)=\frac{1}{\sqrt{(2\pi)^n\det C}}\exp\left[-\frac 1 2 x^t {C}^{-1}x\right],
\label{PxT}
\end{equation}
where $x^t=(x_1,\ldots,x_n)\in\RR^n $ with $t$ denoting the transposition and we have also assumed, without loss of generality, that all the mean values are zero.
Furthermore, $\theta^t=(\theta^1,\ldots\theta^m)$ are the real valued parameters characterizing the above probability distribution function, i.e. the entries of the covariance matrix $C$. Hence $m=\frac{n(n+1)}{2}$.  

Next we consider  a family $\cal P$ of such probability distributions 
\begin{equation*}
{\cal P}=\{p_\theta=p(x;\theta)|\theta^{t}=(\theta^1,\ldots\theta^m)\in\Theta\},
\end{equation*}
where $\Theta\subseteq\RR^m$ and the mapping $\theta\rightarrow p_\theta$ is injective. Defined in such a way $\cal P$ is an $m$-dimensional statistical model on $\RR^n$. The open set $\Theta$ is defined as follows
\begin{equation}\label{parameterspace}
\Theta=\{\theta\in \RR^m| \, C(\theta)>0 \},
\end{equation}
and we call it the parameter space of the $m$-dimensional statistical model $\cal P$.

Assuming parametrizations which are $C^\infty$ we can turn  $\cal P$ into a $C^\infty$ differentiable manifold  \cite{AN00}. Then, given a point $\theta$, the Fisher information matrix of $\cal P$ in $\theta$ is the $m\times m$ matrix $G(\theta)=[g_{\mu\nu}]$, where the $\mu,\nu$ entry is defined by
\begin{equation}
g_{\mu\nu}(\theta):=\int_{\RR^n} dx \;p(x;\theta)\partial_\mu\log p(x;\theta)\partial_\nu\log p(x;\theta),
\label{gFR}
\end{equation}
with $\partial_\mu$ standing for $\frac{\partial}{\partial\theta^\mu}$. The matrix $G(\theta)$ is symmetric, positive semidefinite and endows the parameter space $\Theta$ with a Riemannian metric  \cite{C09}. 

We highlight that the integral of Eq. \eqref{gFR} with \eqref{PxT} is a Gaussian one and amounts to 
\begin{eqnarray}
\exp\left[\frac 1 2 \sum_{i,j=1}^n 
c_{ij}\frac{\partial}{\partial x_i}\frac{\partial}{\partial x_j}\right]f_{\mu\nu} |_{x=0},
\label{Gint}
\end{eqnarray}
where the exponential stands for a power series expansion over its argument (the differential operator) and \begin{equation}\label{f} 
f_{\mu\nu}:=\partial_\mu \log[p(x;\theta)]\ \partial_\nu \log[p(x;\theta)].
\end{equation}
The derivative of the logarithm has the following expression
\begin{equation}\label{logder}
\partial_\mu \log[p(x;\theta)]=-\frac 1 2\Bigg[\frac{\partial_\mu(\det C)}{\det C}
+\sum_{\alpha,\beta=1}^n \partial_\mu(c_{\alpha\beta}^{-1})x_\alpha x_\beta\Bigg],
\end{equation}
where $c_{\alpha\beta}^{-1}$ denotes the $\alpha \beta$ entry of the inverse of the covariance matrix $C$ in \eqref{PxT}.
The latter equation together with Eq. \eqref{Gint} show the computational complexity of the Eq. \eqref{gFR}. 
Indeed, the well-known formulas
\begin{eqnarray*}
\partial_\mu C^{-1}(\theta)&=&C^{-1}(\theta)\big(\partial_\mu C(\theta)\big)C^{-1}(\theta)\\
\partial_\mu(\det C(\theta))&=&\det C(\theta) \;\mbox{Tr}(C(\theta)\,\partial_\mu(C(\theta)))
\end{eqnarray*}
 require the calculation of $n(n+1)$ derivatives with respect to the variables $\theta\in\Theta$ of  Eq. \eqref{parameterspace} in order to work out the derivative of the logarithm in \eqref{logder}. Finally, to obtain the function $f_{\mu\nu}$ in \eqref{f}, we have to evaluate $O(n^4)$ derivatives. This quickly becomes an unfeasible task with growing $n$, even numerically. 

In order to overcome the difficulty of explicitly computing the components of the Fisher-Rao metric, we proceed by defining a new (pseudo)-Riemannian metric on the parameter space $\Theta$  which account as well for the network structure given by the adjacency matrix $A$.

 To this end we consider first a trivial network with null adjacency matrix { and interpret $n$ independent Gaussian random variables $X_i$ as sitting on its vertices. In this particular case, the joint probability distribution \eqref{PxT} is given with a diagonal covariance matrix with entries given by $\theta^i:=\mathbb{E}(X_i^2)$. Let us denote this matrix as $C_0(\theta)$.  }  So, making use of Eqs. \eqref{parameterspace} and \eqref{gFR},  the statistical Riemannian manifold $\mathcal{M}=(\Theta,g)$, with {\cite{jmp}}
 \begin{equation}\label{diagonal}
{ \Theta=\{\theta=\left(\theta^1,\ldots,\theta^n\right)|\theta^i>0\},\ g=\frac{1}{2}\sum_{i=1}^n\Big(\frac{1}{\theta^i}\Big)^2d\theta^i\otimes d\theta^i}
 \end{equation}
is associated to the { \textit{bare}} network.

{Given the matrix $C_0(\theta)=\mbox{diag}\left[\theta^1,\ldots,\theta^n\right]$, from \eqref{diagonal} it is evident that $g_{ii}=\frac{1}{2}(c_{ii}^{-1})^2$, where $c_{ii}^{-1}$ is the $ii$ entry of the inverse matrix of $C_0(\theta)$ given by $c_{ii}^{-1}=\frac{1}{\theta^i}$, for all $i\in\{1,\ldots,n\}$.} Inspired by
this functional form of $g$, we propose  to associate a (pseudo)-Riemannian manifold to any network $\cal X$ with non vanishing  adjacency matrix $A$. { To this aim, we consider the map
 $\psi_{C_0}:\mbox{A}(n,\RR)\rightarrow\mbox{GL}(n,\RR)$ defined by
\begin{equation}\label{psi} 
\psi_{C_0(\theta)}(A):=C_0(\theta)+A,
\end{equation}
with $\mbox{A}(n,\RR)$ denoting the set of the symmetric $n\times n$ matrices over $\RR$  with vanishing diagonal elements that can represent any simple undirected graph. Then, we deform the manifold ${\cal M}$ in \eqref{diagonal} via  $\psi_{C_0}$.}
Hence the  manifold associated to a network $\cal X$ with adjacency matrix $A$ is  $\widetilde{\cal M}=(\widetilde{\Theta},\widetilde{g})$ with 
 \begin{equation}
 \label{varyspace}
 \widetilde{\Theta}:=\{\theta\in\Theta\ \vert\ \psi_{C_0(\theta)}(A)\ \mbox{is non-degenerate}\}
 \end{equation}
 and  $\widetilde{g}=\sum_{\mu\nu}\widetilde{g}_{\mu\nu}d\theta^{\mu}\otimes d\theta^{\nu}$ with components
\begin{equation}\label{gvary}
\widetilde{g}_{\mu\nu}=\frac{1}{2}(\psi_{C_0(\theta)}(A)^{-1}_{\mu\nu})^2,
\end{equation}
where $\psi_{C_0(\theta)}(A)^{-1}_{\mu\nu}$ is the $\mu\nu$ entry of the inverse 
of the matrix $\psi_{C_0(\theta)}(A)$. 

{In this way, we associated a differentiable system (Riemannian manifold) to a discrete system (network) through the description of network by a set of probability distribution functions. Some other ways to describe a network with probabilistic methods are also employed in literature. Among them it is worth mentioning the random walk method \cite{Noh}. Here the Green function, meaning the transition amplitude from one vertex to another by accounting for all possible paths, gives rise to a metric \cite{Blachere}, thus allowing as well for a geometric approach. However the main difference is that in such a case one deals with a stationary transition probability originating from the adjacency matrix \cite{Noh}, in our case beside adjacency matrix also the variances of a Gaussian distribution of random variables sitting on nodes of the network  play a role (namely a family of Gaussian distributions).}

%%%%%%%%%%%%%%%%%%%%%%%%%%%%%%%%%%%%%%%%%%%%%%%%%%%%%%%

\section{A geometric entropy}

We now define a geometric entropy of a network $\cal X$ with adjacency matrix $A$ and associated manifold $\widetilde{\cal M}=(\widetilde{\Theta},\widetilde{g})$ as
\begin{equation}
{\cal S}:= \ln {\cal V}(A),
\label{entropy}
\end{equation}
where ${\cal V}(A)$ is the volume of $\widetilde{\cal M}$ evaluated from the element 
\begin{equation}
\nu_g=\sqrt{|\det \widetilde{g}(\theta)|}\;d\theta^1\wedge\ldots\wedge d\theta^n\ .
\label{volumelement}
\end{equation}
Notice, however, that in such a way ${\cal V}(A)$ is ill-defined.  Thus we regularize it as follows
\begin{equation}
{\cal V}(A):=\int_{\widetilde{\Theta}}\Upsilon(\psi_{C_0(\theta)}(A))\;\nu_g,
\label{reg}
\end{equation}
where   $\Upsilon(\psi_{C_0(\theta)}(A))$ is any suitable  "infrared" and "ultraviolet" regularizing function;
$\psi_{C_0(\theta)}(A)$ and $\nu_g$ are  given in \eqref{psi} and \eqref{volumelement}, respectively. The need for regularization is twofold: the set $\widetilde{\Theta}$  in Eq.\eqref{varyspace} is not compact because  the variables $\theta^i$ are unbounded from above; furthermore, from Eq.\eqref{gvary}, $\det \widetilde{g}(\theta)$ diverges
since $\det \psi_{C_0(\theta)}(A)$ approaches zero for some $\theta^i$. { A possible choice of $\Upsilon$ has recently been defined \cite{jmp},
\begin{equation}\label{regjmp}
\Upsilon(C(\theta)):= e^{-\mbox{Tr} \left(C(\theta)\right)}\ \log\left[1+\left(\det C(\theta)\right)^n\right],
\end{equation}  
where $C$ is the covariance matrix in \eqref{PxT} when off-diagonal entries are $1$ or $0$. In the present work we would extent such a regularizing function to a more general kind of networks, taking also into account the weights of links between vertices. However, the functional type should be still like in \eqref{regjmp}.}

The definition \eqref{entropy} is inspired by the microcanonical definition of entropy $S$ in Statistical Mechanics, that is $S:=k_B \ln \Omega(E)$, where $\Omega(E)$ is the phase space volume bounded by the hypersurface of constant energy $E$. After integration on the momenta one finds $S = k_B \ln  \varpi \int_{M_E} [E - V(q_1,\ldots,q_n)]^{n/2}\;dq^1\wedge\ldots\wedge dq^n $, where $ {\varpi}$ is a constant stemming from the integration on the momenta, $M_E$ is the configuration space subset bounded by the equipotential level set $E=V(q_1,\ldots,q_n)$, and $q_1,\ldots, q_n$ are the configurational coordinates. Now, the term $ [E - V(q_1,\ldots,q_n)]^{n/2}$ is just $\sqrt{\det g_J}$, with $g_J$ the Jacobi kinetic energy metric whose associated geodesic flow coincides with the underlying Hamiltonian flow \cite{P07}. In the end the microcanonical entropy is $S=k_B \ln \int_{M_E} \sqrt{\det g_J} \;dq^1\wedge\ldots\wedge dq^n + k_B \ln {\varpi}$, that is proportional to the logarithm of the volume of the Riemannian manifold associated with the underlying dynamics.

As already stated at the beginning of this paper, in order to assess the interest of the proposed geometric entropy in Eq.\eqref{entropy} we check it against a system undergoing the classical Erd\"{o}s-R\'enyi phase transition in random graphs \cite{ER,Lucz}.

%%%%%%%%%%%%%%%%%%%%%%%%%%%%%%%%%%%%%%%%%%%%%%%%%%%%%%%

\section{Numerical results}

We numerically  compute ${\cal S}(k)$, the geometric entropy in Eq.\eqref{entropy} vs $k$ for a fixed $n$ in order to investigate its sensitivity to the appearance of the giant component during the evolution of the random graph model $\mathbb{G}(n,k)$. 

To this aim, we consider four different numbers of vertices:  $n=25, 50, 100, 200$. 
The choice of $n$ determines the dimension of the associated manifold ${\widetilde{\cal M}}$. Then, for a given $n$, we choose the number of links $k$, with $k=0,1,\ldots,n(n-1)/2$.   Next, for a given pair $(n,k)$ we generate at random a set of $k$ entries $(i,j)$, with $i<j$, of the non-vanishing adjacency matrix elements $A_{ij}$.

\begin{figure}\centering
\includegraphics[width=8cm,height=6cm,scale=1.]{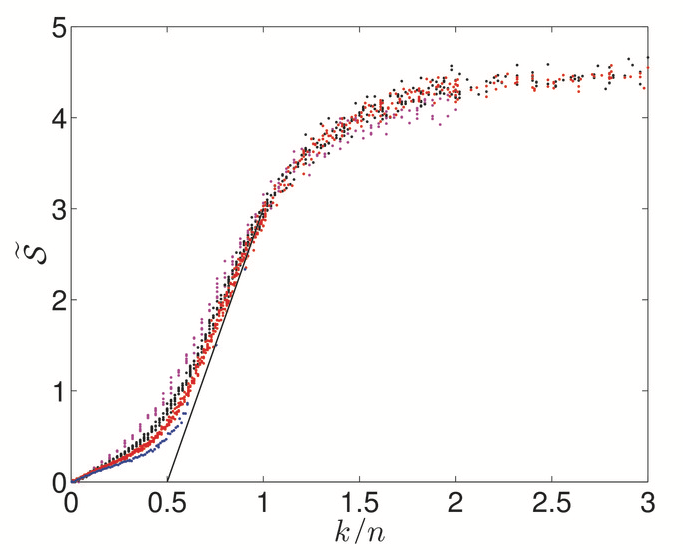}
\caption[10 pt]{(Color on line) Geometric entropy ${\widetilde{\cal S}}$ of  $\mathbb{G}(25,k)$ (magenta points), $\mathbb{G}(50,k)$ (black points), $\mathbb{G}(100,k)$  (red points) and $\mathbb{G}(200,k)$  (blue points) networks as a function of the number $k$ of randomly chosen links of weights equal to $r=0.2$. The black solid line is a guide to the eye coming from a linear fitting of a linear-logarithmic presentation of the data.}
\label{bifurcation}
\end{figure}

\begin{figure}\centering
\includegraphics[width=8cm,height=6cm,scale=1.]{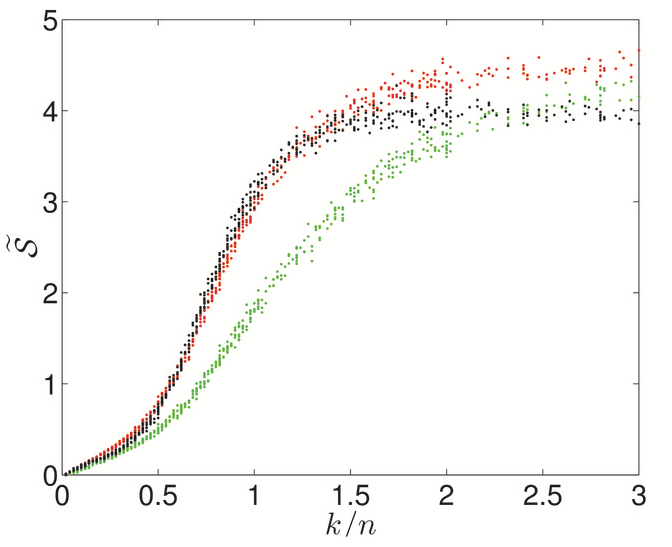}
\caption[10 pt]{(Color on line) Geometric entropy ${\widetilde{\cal S}}$ of  $\mathbb{G}(50,k)$ networks as a function of the number $k$ of randomly chosen links of weights equal to  $r=0.1$ (green points), $r=0.2$ (red points) and $r=0.4$ (black points).}
\label{linkweight}
\end{figure}

Hence, since the covariance  matrix $C$ is functionally assigned, we get $\psi_{C}(A)$ of  Eq. \eqref{psi} and finally the metric $\widetilde{g}$ of Eq.\eqref{gvary}. Now, having determined ${\widetilde{\cal M}}=({\widetilde{\Theta}},\widetilde{g})$, we can compute the volume ${\cal V}(A)$ in Eq.\eqref{reg} and the entropy ${\cal S}$ of Eq.\eqref{entropy}. In numerical computations the volume regularization  is performed in two steps, the first one is by restricting the manifold support ${\widetilde{\Theta}}\subset \mathbb{R}^n$ to an hypercube.
Inside ${\widetilde{\Theta}}$ we generate a Markov chain, to perform a Monte Carlo estimate of the average 
\begin{equation*}
\left\langle\sqrt{\det \widetilde{g}} \right\rangle =\frac{\int \sqrt{\det \widetilde{g}}\;d\theta^1\wedge\ldots\wedge d\theta^n}{\int \;d\theta^1\wedge\ldots\wedge d\theta^n}.
\end{equation*}
The number of random configurations considered varies between $10^4$ and $10^6$; the second step of the regularization procedure of the volume is obtained by excluding those points where the value of $\sqrt{\det \widetilde{g}}$ exceeds $10^{308}$ (the numerical overflow limit of the computers used). For any given pair $(n,k)$ this computational scheme is repeated $10^3$ times, each time considering a different randomly generated realisation of the adjacency matrix $A$. Thus the final values of the entropy  ${\cal S}$ are obtained as averages on these $10^3$ different manifolds ${\widetilde{\cal M}}$. In the Figures \ref{bifurcation} and \ref{linkweight} we report the Monte Carlo numerical estimates of
\begin{eqnarray}
{\widetilde{\cal S}}(k) &:=& \frac{\langle ({\cal S}(k) - {\cal S}(0))\rangle}{n}\nonumber \\
&=& \frac{1}{n} \left\langle\ln \frac{\int \sqrt{\det \widetilde{g}}\;d\theta^1\wedge\ldots\wedge d\theta^n}{ \int \sqrt{\det { g}} \;d\theta^1\wedge\ldots\wedge d\theta^n}\right\rangle
\label{collapseplot}
\end{eqnarray}
repeated for different values of $k$; { here} $\langle\cdot\rangle$ stands for the above mentioned average  over  the different realisations of the adjacency matrix $A$, and { $g$} is the metric corresponding to the null adjacency matrix.

For all the four cases reported in Figure \ref{bifurcation} we have equal weights { $A_{ij}=r$} for all the $k$ non-vanishing links.

The reason for displaying ${\widetilde{\cal S}}$ of Eq.\eqref{collapseplot} instead of ${\cal S}$ in \eqref{entropy} and \textit{versus}  $k/n$ instead of $k$ is that one obtains what in the context of statistical mechanics is called a \textit{collapse  plot} of the results obtained at different $n$-values. The corresponding points crowd on a common pattern for large $k$ whereas for $k/n$ ranging from $0$ to approximately $1$ the patterns obtained show a phenomenon which is familiar in the context of numerical investigations of second order phase transitions: as in the case of finite-size effects observed for the order parameter, what asymptotically would be a sharp bifurcation is rounded at finite $n$ \cite{P07}; however, the larger $n$ the more pronouced the `knee'  of ${\widetilde{\cal S}(k/n)}$ in the range $(0,1)$. This is clearly in excellent agreement  with an $n$-asymptotic bifurcation at $k/n = 0.5$ (marked by the solid line) where the Erd\"{o}s-R\'enyi phase transition takes place. 

In Figure \ref{linkweight} we report the outcomes for $\mathbb{G}(50, k)$ having set all the non-vanishing entries $A_{ij}$ of the adjacency matrix { again equal to a constant value $r$.} For $r=0.1$ a considerable softening of the shape of ${\widetilde{\cal S}}(k/n)$ is observed; this is of course an expected result since for $r\to 0$ the transition must disappear. For $r=0.2$ and $r=0.4$ the shapes of ${\widetilde{\cal S}}(k/n)$ look almost the same, the only interesting difference being a slightly more pronounced `knee' in the $r=0.4$ case, which, going in the opposite direction, is coherent with the previous ones.

Another interesting property of this entropy consists in its versatility, that is, it can be easily adapted to more refined descriptions of networks/graphs. 

For example, we can consider a refined modeling of graphs where the entries of the $n\times n$ adjacency matrix $A$
are given by terms of the form $r_{ij} \theta^i \theta^j$. Here the $r_{ij}$s ($i,j=1,\ldots,n$) are the weights of the links between the nodes of the network described by $A$.
Furthermore, the $\theta^i$s ($i=1,\ldots,n$) are local coordinates on the manifold 
$\widetilde{\cal M}$ of
{Eqs.\eqref{varyspace} and \eqref{gvary},} representing the variances of the random variables on the nodes of the network.

\begin{figure}\centering
\includegraphics[scale=0.32]{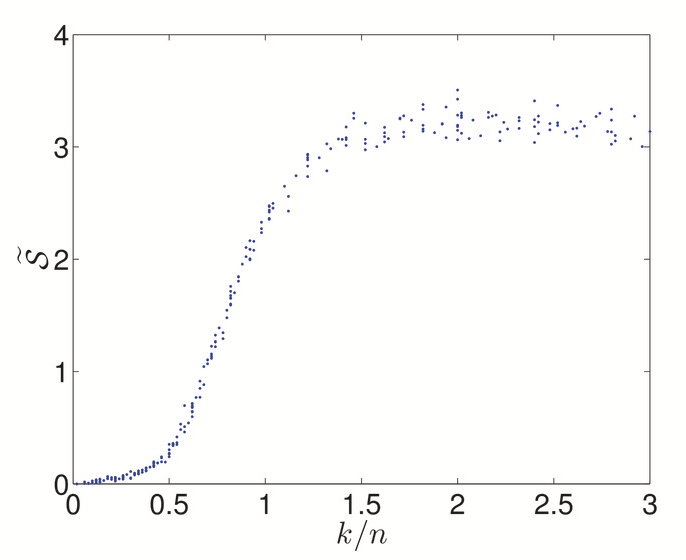}
\caption[10 pt]{ (Color on line) Geometric entropy of $\mathbb{G}(50,k)$ networks as a function of the number $k$ of randomly chosen non-vanishing adjacency matrix elements $A_{ij}$ of the form $r_{ij} \theta^i \theta^j$ with $r_{ij}=0.2$.}
\label{linkthetaithetaj}
\end{figure}

This kind of model has an interesting property: the closer a given variable $\theta^i$ to zero, the weaker the  weights of all the links associated to the $i$-th node. In such a way, this second model, allows one to
describe a more general class of networks. In fact, consider for example the flow of some quantity across a network,  the vanishing of the flow on a given node $i$ implies that all the other nodes connected to it become effectively independent of it. In view of this argument, we consider the entropy $\cal S$ in Eq. \eqref{entropy} against the more general
model just described above. 
In Figure \ref{linkthetaithetaj} it is shown ${\widetilde{\cal S}}$ versus $k/n$ where $n=50$ is the dimension of the manifold
associated to the network, and $k=0,\ldots, n(n-1)/2$ is the number of non-vanishing $r_{ij}$ (all of them are chosen equal
to $1$), for $i,j = 1,\ldots,n$ and $i<j$. Also in this case our entropy detects the phase-transition
predicted by the Erd\"{o}s-R\'enyi theorem occuring at $k/n=0.5$. The pattern of  ${\widetilde{\cal S}}(k/n)$ reported in Figure \ref{linkthetaithetaj} shows a more pronounced "knee" at the asymptotic transition value $k/n=0.5$ with 
respect to what is found for the same $n$ value and is reported in Figures \ref{bifurcation} and \ref{linkweight}.

%%%%%%%%%%%%%%%%%%%%%%%%%%%%%%%%%%%%%%%%%%%%%%%%%%%%%%

\section{Conclusion}

Summarizing, the present work puts forward a novel entropic functional useful to characterize probabilistic graph models.
It is inspired to Statistical Mechanics, however, instead of being modeled on the Boltzmann entropy it is rather modeled on the microcanonical ensemble definition of entropy. The phase space volume being replaced by the volume of a `state manifold' (that is a Riemannian manifold whose points correspond to all the possible states of a given network). The state manifold is defined through a suitable metric which is borrowed from an information geometry framework. The result is a constructive way of associating a differentiable and handy mathematical object to any simple undirected and weighted graph or network. 

Notice that a similar way of associating a probability distribution to a network, is that of probabilistic graphs models \cite{K12}.  Here the choice of Gaussian probability distributions is motivated by the fact that Gaussian networks are extensively used in many applications ranging from neural networks, to wireless communication, from proteins to electronic circuits, and so on.

The most relevant property of the proposed entropic-geometric measure is its ability to 
detect the phase transition which is rigorously predicted by the Erd\"{o}s-R\'enyi theorem for random graphs:  a paradigmatic example of an analytically known emergent phenomenon occurring in a network. 
This effect shows up very clearly, in fact,  the geometric entropy proposed here displays both the pattern (as a function of a control parameter) and the finite-size-dependence which are typically displayed by the order parameter of a second-order phase transition in physics.
As natural, though non-trivial, extension our entropic-geometric measure could be applied to pseudo-random graphs and dense graphs where the emergence of the giant component has been recently proved \cite{Frieze,Bollobas}.

Finally, the differential-geometric framework proposed opens some fascinating perspectives of application to the study of complex networks \cite{CNet2,CNet2bis}. 
As matter of fact the introduced geometric entropic measure could account for both the structural complexity of a given network and for its statistical complexity, that is, for the complexity of the probability distributions of the entities constituting the network.

%%%%%%%%%%%%%%%%%%%%%%%%%%%%%%%%%%%%%%%%

%%%%%%%%%%%%%%%%%%%%%%%%%%%%%%%%%%%%%%%%%%%%%%%%%%%%%%

\section*{Acknowledgments}
We acknowledge the financial support of the European Commission by the FET-Open grant agreement TOPDRIM, number FP7-ICT-318121.

\end{document}